\RequirePackage[columnwise]{lineno}
\documentclass[reprint,aps,amsmath,amssymb,notitlepage,superscriptaddress, twocolumn,color,epsfig,graphicx,bm,floatfix,nofootinbib,longbibliography]{revtex4-1}
\usepackage[utf8]{inputenc}

\usepackage{lipsum}
\usepackage{tikz}

\usetikzlibrary{decorations.markings}

\usepackage{amsmath}  
\usepackage{url}
\usepackage{hyperref}
\usepackage{mathtools}
\usepackage{siunitx}
\usepackage{multirow}
\usepackage{microtype}
\usepackage{array}
\usepackage{tabulary}
\usepackage{longtable}
\usepackage[version=4]{mhchem}
\usepackage{graphicx}
\newcolumntype{K}[1]{>{\centering\arraybackslash}p{#1}}

\usepackage{graphicx}   
\usepackage{verbatim}   
\usepackage{color}      
\usepackage{subfigure}  
\usepackage{hyperref}   

\newcommand{\s}{$\langle\sigma\rangle$}
\newcommand{\sigsig}{$\langle\sigma\sigma\rangle$}
\newcommand{\Sss}{$\langle\sigma\rangle\langle\sigma\sigma\rangle$}
\usepackage{lipsum}

\graphicspath{{graphics/}}

\begin{document}

\title{Transfer learning  in predicting quantum many-body dynamics: from physical observables to entanglement entropy}

\author {Philipp Schmidt}
\affiliation {Max-Planck-Institut f{\"u}r die Physik des Lichts, Staudtstrasse 2, 91058 Erlangen, Germany}

\affiliation{Physics Department, University of Erlangen-Nuremberg, Staudtstr. 5, 91058 Erlangen, Germany}

\author{Florian Marquardt}
\affiliation {Max-Planck-Institut f{\"u}r die Physik des Lichts, Staudtstrasse 2, 91058 Erlangen, Germany}

\author {Naeimeh Mohseni}
\affiliation {Max-Planck-Institut f{\"u}r die Physik des Lichts, Staudtstrasse 2, 91058 Erlangen, Germany}

\affiliation{Physics Department, University of Erlangen-Nuremberg, Staudtstr. 5, 91058 Erlangen, Germany}

\date{\today}
\begin{abstract}

Deep neural networks have demonstrated remarkable efficacy in extracting meaningful representations from complex datasets. This  has propelled representation learning as a compelling area of research across diverse fields. One interesting open question is how beneficial representation learning can be for quantum many-body physics, with its notouriosly high-dimensional state space. In this work, we showcase the capacity of a neural network that was trained on a subset of physical observables of a many-body system to partially acquire an  implicit representation of the wave function. We illustrate this by demonstrating the effectiveness of reusing the representation learned by the neural network to enhance the learning process of another quantity derived from the quantum state. 
In particular, we focus on how the pre-trained neural network  can enhance the learning of entanglement entropy. This is of particular interest as directly measuring the entanglement in a many-body system is very challenging, while a subset of physical observables can be easily measured in experiments. We show the pre-trained neural network learns the dynamics of entropy with fewer resources and higher precision in comparison with direct training on the entanglement entropy.

\end{abstract}
\maketitle
\section{Introduction\label{sec:introduction}}

Representation learning has emerged as a captivating avenue within machine learning research \cite{bengio2013representation}.  
It has created significant advancements in various fields such as language modeling \cite{devlin2018bert,bowman2015generating} and computer vision \cite{karras2017progressive,krizhevsky2017imagenet}, revolutionizing the way data is understood and processed. 
Investigating the learned representation by a machine learning model provides insights into which features of the model are most relevant. This offers a fresh perspective on the model's understanding, potentially revealing nuances beyond human perception.  In physics, especially in quantum many-body systems, representation learning provides valuable insight into the essential properties of complex quantum systems.  For example, this technique has successfully been used to create compressed representations of quantum states \cite{rocchetto2018learning, sun2024quantum, schmale2022efficient}.

In this work, which is geared towards applying representation learning in quantum many-body physics, we investigate the capability of a neural network pre-trained on the prediction of physical observables to build up (partially) an implicit representation of the complex many-body wave function. To demonstrate this potential, we explore how the pre-trained neural network enhances the learning process of other tasks that depend on the quantum state.

As an example of a task depending on a suitable representation of the quantum state,  we focus on the prediction of entanglement between parts of the many-body system. Entanglement is  a fundamental concept in quantum mechanics  with wide-ranging implications for  fundamental physics \cite {aspect1999bell} and practical applications in quantum computing and quantum communication \cite{RevModPhys.80.517, liao2017satellite, PhysRevA.101.013804}.  Efficiently detecting and quantifying the entanglement of a quantum many-body system is a very challenging task.  The conventional methods for entanglement detection necessitate demanding full quantum state tomography \cite{RevModPhys.84.777}. Nevertheless, alternative approaches have emerged that alleviate this requirement. For example, there exist methods based on partial state tomography \cite{peres1996separability} or reduced density matrices \cite{emonts2022reduced}, which only require measurements of a subset of the degrees of freedom. 
Meanwhile, machine learning techniques have enhanced  entanglement detection schemes \cite{gao2018experimental, qiu2019detecting, ma2018transforming,harney2020entanglement, asif2023entanglement}. 
The key approach in all these works is to apply classifiers to detect entanglement based on certain features, without requiring full information about the wave function \cite{ qiu2019detecting, ma2018transforming,harney2020entanglement}.

\begin{figure}[t]
\centering
\includegraphics[width=1\linewidth]{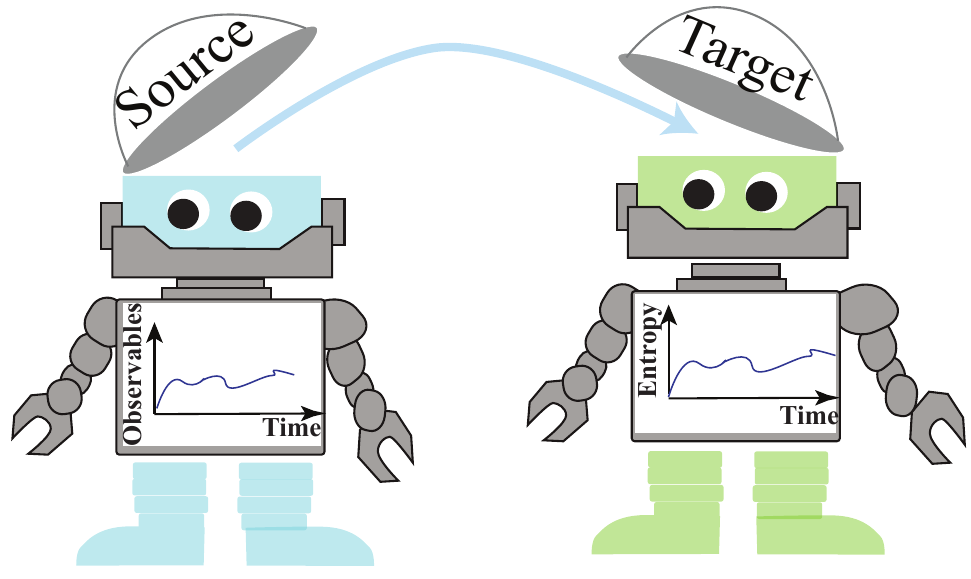}
\caption {Schematic illustration of the transfer learning task considered in this study. The information learned by a neural network trained on the dynamics of a subset of physical observables of  a many-body system is reused to learn efficiently  the dynamics of entanglement entropy. 
 \label{fig1_1}}
\end{figure}

\begin{figure}[t]
\centering
\includegraphics[width=1\linewidth]{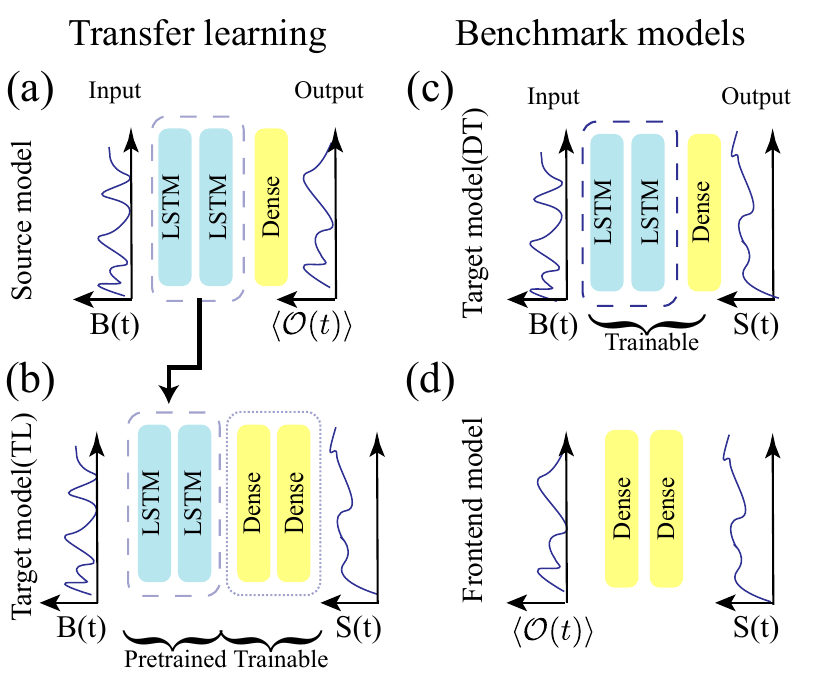}
\caption {Schematic illustration of the models employed in this study. $B(t)$, $\langle \mathcal{O}(t) \rangle$, and $S(t)$ represent the magnetic field, a  subset of physical observables, and    von Neumann entropy, respectively. (a) Source model; the neural network is fed by the magnetic field at each time  step and it outputs a subset of physical observables for the corresponding time step. (b) The target model with transfer learning (TL) which  is built of two parts. The first component comprises layers that have been pre-trained on the dynamics of physical observables within the source model. These pre-trained layers remain fixed during the training process of the target model. They are then stacked with  extra layers which are trainable. (c)  Direct training (DT) of the target model; all the layers are trainable and no prior information is passed to this architecture. (d) In the front-end model, the neural network receives as input the time evolution of a subset of physical observables and outputs the time evolution of entropy. Note that in the source model and the target models, the initial values of $\langle\sigma_i^\alpha \rangle$ are also fed as input to the neural networks. 
 \label{fig1}}
\end{figure}

Motivated by the interest in learning entanglement, we inspect  whether our  data-driven neural network that has been trained on predicting the dynamics of a subset of  physical observables of a quantum  many-body system  can be reused to learn the dynamics of entanglement with fewer resources compared to the direct training on entanglement entropy.  
    Such a technique  where a model implemented for a particular task (called source model) can be reused as the starting point for modeling  a second task (called  target model) is a prevailing tool in deep learning  and is known as transfer learning (TL) \cite{yang2008introduction,caruana1998multitask}.


 We examine the efficiency of a pre-trained neural network on physical observables in learning the dynamics of entanglement entropy for both integrable and non-integrable models. Notably, we demonstrate that in integrable models, where the network exhibits superior capability in learning the physical observables \cite{mohseni2022deep, mohseni2023deep}, the accuracy of transfer-learning-based predictions for the target model is also  higher.

\section{Physical Model}
We conduct our experiments on two  typical  spin models.   An Ising ring  driven with a time-dependent transverse magnetic field in the presence or absence of an extra (integrability-breaking) longitudinal field is described by the following Hamiltonian:

 \begin{equation}
        H = B(t)\sum_{i=1}^{n}\sigma_i^x + J\sum_{i=1}^{n}\sigma_i^z\sigma_{i+1}^z+g \sum_{i=1}^{n}\sigma_i^z
        \label{eqn:TSI}
    \end{equation}

Our choice is motivated by the fact that while the case of $g=0$ is instantaneously (at any fixed time)  quantum integrable,  in the presence of the extra longitudinal field the model  is non-integrable.    The time-dependent random trajectories for the magnetic field are generated using a random Gaussian process \cite{liu2019advances}; see Supplemental Material in Ref. \cite{mohseni2022deep} for technical details.  Note that  we consider closed boundary conditions so that $\sigma^\alpha_{n+1} \coloneqq \sigma^\alpha_{1}$. We use qutip \cite{johansson2012qutip} to calculate the von Neumann entropy of the reduced density matrix $\rho_r$ of a subsystem, defined as $S=-T r (\rho_r \ln \rho_r)$, for the different realizations of our time-dependent random magnetic field.  

\section{Methodology}
    \label{sec:methods}
    \paragraph{\textbf{Neural network architecture}}
We aim to investigate the efficiency of a pre-trained neural network on physical observables, in learning the  dynamics of entanglement entropy.  To explore our objective, we train four distinct models:  A source model, a transfer learning of the target model, a direct training of the target model as well as a front-end model.  The source model refers to the neural network that is fed with the value of the magnetic field at each time step as well as the initial values of $\langle\sigma_i^\alpha \rangle$ and outputs the desired physical observables for that time step (Fig. \ref{fig1} (a)).  The target model is the neural network architecture that receives as  input the magnetic field at each time step as well as the initial values of $\langle\sigma_i^\alpha \rangle$ and outputs the von Neumann entropy for the corresponding time step.  For the direct training (DT) of the target model, all the layers are trainable and no prior information is passed to this architecture Fig. \ref{fig1} (c).  In contrast,  the target model with transfer learning (TL) is made of two parts: The first part is represented by layers that are  pre-trained on the dynamics of physical observables in the source model and are frozen during the training of the target model. These pre-trained layers are stacked with  extra layers that are trainable (Fig. \ref{fig1} (b)), which constitute the second part. 
 
 As we explain later in detail, to gain further insights into the significance of trainable layers and the utility of information derived from pre-trained layers in transfer learning of the target model we also train another model referred to as the front-end model. In this model, the network receives as input the dynamics of physical observables and it outputs  the dynamics of entropy as shown in Fig. \ref{fig1} (d).

Our selection of layer structures for each model is based on the following considerations.  For the task of training the source model on a subset of physical observables, consistent with our previous studies \cite{nmohseni2021deep, mohseni2022deep, mohseni2023deep}, our observations suggest that Long Short-Term Memory (LSTM) layers offer improved performance.  The same observation holds for the direct training of the target model. Therefore, we choose LSTM layers for both the source model and direct training of the target model.  Regarding transfer learning of the target model, the process of transferring information from a source model to a target model requires smart fine-tuning. To transfer layers from a source model to a target model successfully, it's crucial to choose layers of the source model that retain the most important information. Our experiments indicate that transferring all hidden layers of the source model is beneficial ( see Appendix Sec. \ref{app:additional} and Fig.  \ref{fig:appendix} (c) for a detailed discussion). We observed that selecting a dense layer as the output layer of the source model   leads to \textit{slightly} better performance in the source model itself (see  Appendix Sec. \ref{app:additional} and Fig.  \ref{fig:appendix} (a)) as well as  a \textit{slightly} more efficient transfer of information to the target model (see Appendix Sec. \ref{app:additional}  and Fig.  \ref{fig:appendix} (a)). However, it is worth noting that the observed differences are not substantial in magnitude.

In the target model with transfer learning, the trainable layers should be selected in a manner that enables them to effectively extract the information from the frozen layers.   It is  also important to compare the power of the extra trainable layers  with the architecture chosen for direct training.  Our observations (see Fig. \ref{fig:appendix} (b)) in general  indicate that LSTM layers are more resource-efficient in comparison with dense layers in extracting information from pre-trained LSTM layers. However, it is possible that the better performance  in comparison to direct training could be solely attributed to the trainable part. To validate that the information from the pre-trained layers also plays a role, we consider the scenario where the trainable layers are  dense and therefore less powerful. This allows us to further confirm the contribution of the pre-trained layers to the performance of the target model. All the results of the paper are produced with the dense trainable layers, i.e. the less powerful layout. 

For the front-end model, LSTM layers would typically be more suitable. However, we have chosen to use dense layers instead. This choice allows us to maintain a comparable level of power between this model and the trainable layers of the target model during transfer learning. By doing so, we ensure a fair and meaningful comparison between the models.  
 See Appendix Sec. \ref{app:tech_details} for more detail on the layout of each model.

\paragraph{\textbf{Training}:} 
 Training and test data for both source and target model are generated by solving the Schrodinger equation for  the Hamiltonian \eqref{eqn:TSI} using qutip \cite{johansson2012qutip}, a library implemented in python.
We prepare our spins initially in an arbitrary translationally-invariant uncorrelated state $\bigotimes_i(\sqrt{p}\vert 0 \rangle+\sqrt{1-p}\vert 1\rangle)$ with $p$ chosen at random from the interval $[0,1]$. Here, $|0\rangle$ and $|1\rangle$ denote, respectively, the $\pm 1$ eigenstates of $\sigma_z$. 

To train our source model we feed to the network as input the time-dependent magnetic field trajectory and the initial values of  $\langle\sigma^\alpha \rangle$ with $ \alpha \in \{x, y, z\}$.   The output of the neural network is the dynamics of a subset of $\langle\sigma_i^\alpha \rangle$  and  $\langle\sigma_i^\alpha \sigma^\beta_{i+\ell} \rangle$ with $ \alpha, \beta\in \{x, y, z\}$. We inspect later how the number of chosen observables to train the source model affects the performance of the target model in learning the entropy.
The cost function that we use to train our source model is defined as
\begin{equation}
{\textrm{MSE}}=\overline{|\langle\mathcal{O}(t)\rangle_{\rm NN}-\langle\mathcal{O}(t)\rangle_{\rm true}|^2} \label{eq2}
\end{equation}
where the average is performed over all samples and time steps. The $\langle\mathcal{O}(t)\rangle$ shows the expectation values of all observables where ``NN" stands for the predictions of the neural network and ``True" stands for the ground-truth values calculated by solving the Schrödinger equation.

To train our target model we feed to the neural network as input the magnetic field and  the initial values of  $\langle\sigma^\alpha \rangle$. The network's output represents the entropy dynamics calculated for the density matrix of one half of the spin ring.
The cost function for the target model is also defined as 
\begin{equation}
\label{eqn:cost_S}
{\textrm{MSE}}=\overline{|S(t)_{\rm NN}-S(t)_{\rm true}|^2} 
\end{equation}
where $S(t)$ represents the evolution of entropy and the average is taken over all samples and time
steps. 
For the front-end model we use the same cost function as for the target model, defined in equation \eqref{eqn:cost_S}.

\section{Results} 
    \label{sec:results}
\begin{figure*}[t!]
            \centering
            \includegraphics[width=\textwidth]{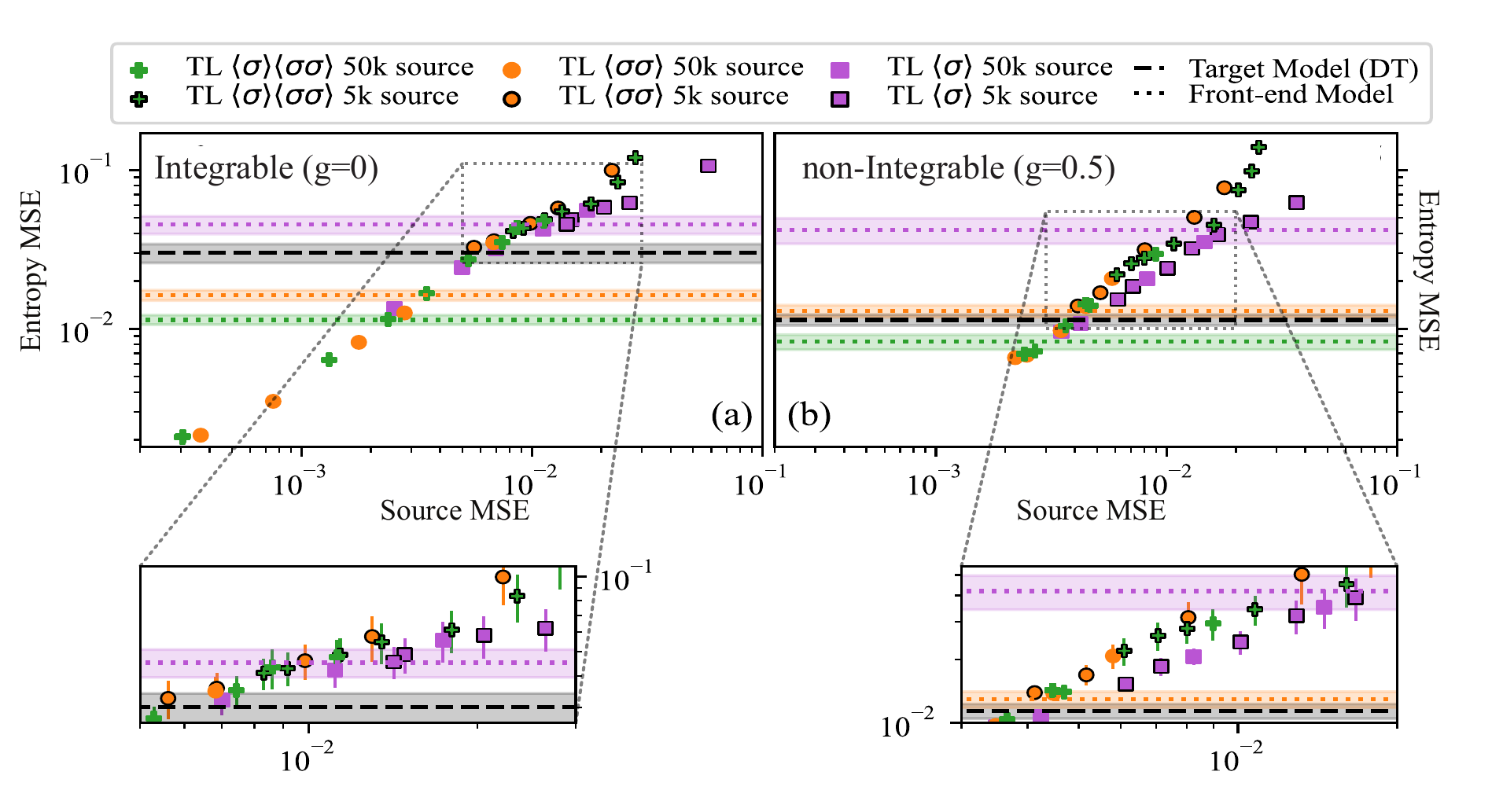}
            \caption{Correlation between the accuracy of the source model and the accuracy of the TL. Results are shown for (a) the integrable model and (b) the non-integrable model for a spin ring with size $N=8$. We consider three scenarios where three different sets of observables are used to train the source model (see Table \ref{tab:obs_sets}). For each set of observables, the source model is trained for two different training set sizes of 5$K$ and 50$K$. The TL model is trained on 5$K$ samples. The horizontal dashed  black line shows the performance for direct training of the target model where 50$K$ samples are used for training. The dotted lines show the MSE for the front-end model where  5$K$  samples are used for training for different sets of physical observables distinguished by color. The colored areas show the variance of the MSE. MSE is always computed over  1000 test samples. The scenarios positioned below the black dashed line denote regions where TL outperforms the direct training of the target model. }
            \label{fig2}
    \end{figure*}
In this section, we primarily aim to investigate the correlation between the accuracy of the transfer learning model in predicting the dynamics of entropy and the accuracy of the source model in learning the physical observables dynamics. Furthermore, we explore how the selection of a subset of observables used to train the source model influences the performance of the target model in the task of transfer learning. To comprehensively assess these aspects, we compare the effectiveness of transfer learning in capturing the evolution of the von Neumann entropy against direct training. Additionally, we compare its effectiveness with the performance of a front-end model that directly maps the dynamics of physical observables to entropy dynamics.


In our previous work \cite{mohseni2022deep}, we observed that the neural network learns the   dynamics of physical observables with higher accuracy for integrable models in comparison with non-integrable models. Therefore, here we explore the performance of the transfer learning of the  target model separately for non-integrable and integrable models. 

Note that we examined the neural network's ability to learn the evolution of bi-partite von Neumann entropy for varying subset sizes of our spin ring and found that the neural network generally performs better for smaller subset sizes (See Appendix Sec. \ref{app:additional} and Fig. \ref{fig:appendix} (d). Our analysis in the main text  is therefore focused only on the most difficult scenario, namely the von Neumann entropy across two halves of a spin ring.

  In Fig. \ref{fig2} (a), for the case of $g=0$, where the model is integrable,  we show the MSE  in the target model defined in Eq. \eqref{eqn:cost_S}  versus MSE in the source model defined in Eq. \eqref{eq2}. The presented results pertain to scenarios where distinct sets of observables are employed to train the source model; see Appendix Sec. \ref{app:tech_details}, Table. \ref{tab:obs_sets} for details on observables labels.  For each set of observables, the source model is trained separately for two different training set sizes of 5$K$ (5,000) and 50$K$. For both cases the number of samples used to train the TL model is 5$K$.   
  
  The correlation between the precision of the source model and the performance of the TL is clearly evident - a higher precision in the source model results in improved TL performance. Additionally, it seems that two-point correlators have a more pronounced impact (compared to the first order moments of spin operators) on the success of the TL. Once a sufficiently high level of accuracy is attained in the source model for each observable set, TL has the potential to outperform its corresponding target model based on direct training and the front-end model  significantly.
  
  We also show the accuracy of the direct training of the target model (dashed line) and the front-end model (which is trained for the different sets of observables) in predicting the dynamics of entropy. As is evident, for cases where the source model learns observables with a higher accuracy, the TL has a better performance.  Note that  a direct comparison of  TL with the front-end model may not be entirely reasonable, as they serve different purposes. The former maps magnetic fields to the evolution of entropy, while the latter maps the evolution of physical observables to the evolution of entropy. However, despite this distinction, we  still draw a comparison as TL employs frozen pre-trained layers from the source model, which map magnetic fields to physical observables. Note that the trainable part of the target model with transferred learning has the same power in terms of neural network architecture and number of trainable layers as the front end model.

  In Fig. \ref{fig2} (b), we show the same plot for the non-integrable case where $g=0.5$.  The difference between direct training and TL is much  smaller in comparison with the integrable case. Training the source model on two-point correlators results in a  slightly better performance of TL in comparison with direct training, which is however not significant. 
   The lower accuracy of TL for the non-integrable model can be attributed to the lower accuracy of the source model.  Additionally only TL based on pre-training first order observables significantly outperforms its corresponding front-end model.   It is important to highlight that even in the scenarios where transfer learning (TL) does not offer a noticeable advantage in terms of prediction accuracy, TL still remains a more resource-efficient approach. This is because TL only uses 5$K$ samples, whereas direct training uses 50$K$ samples for entropy. This implies that, despite potentially comparable prediction accuracy, TL significantly reduces the data requirements to directly train on entropy. We remark that reducing the number of samples to learn entropy  comes at the expense of providing sufficient data for physical observables. These, however, can be measured in experiments directly and easily. We recall  that all the results are for system size $N=8$ and the von Neumann entropy is computed between the two halves of the spin ring.



\begin{figure*}[t]
                \centering
\includegraphics[width=0.9\textwidth]{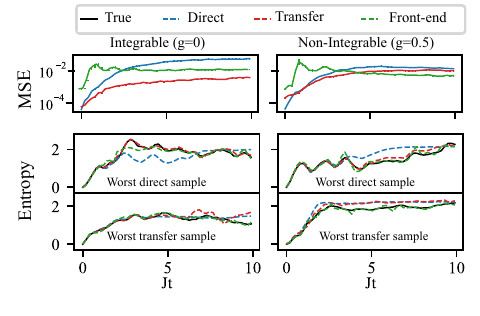}
                \caption{Demonstration of the advantage of TL over direct training  on predicting the dynamics of entropy for an integrable model (g=0, left column) versus a non-integrable model (g=0.5, right column) for system size $N=8$.    The first row shows the time evolution of MSE (defined in Eq. \eqref{eqn:cost_S}) averaged over 1000 test samples. For the integrable model, it is evident that the TL model predicts entropy with higher accuracy in comparison with both direct training and the front-end model. For the non-integrable model, such difference is not evident which is due to the low accuracy of the source model. The second and third rows  show the cases for which direct training and transfer learning performed worst out of 1000 test samples, respectively.}
                \label{fig3}
            \end{figure*}

In Fig. \ref{fig3}, we demonstrate further  the utility of the source model in scenarios where the target model lacks sufficient data.  We compare the accuracy of  predicting the dynamics of entropy applying the TL of the target model with the direct training of the target model as well as with the front-end model. Here the training set size on  the source model, direct training and   transfer learning of the target model as well as the front-end model are, 50$k$, 50$k$, 5$k$ and 5$k$, respectively.  The left  column (right)  shows the results for the integrable model (non-integrable).  Out of 1000 test samples, the first row represents the true and predicted entropy evolution for a single instance  where direct training performs worst and the second panel presents the case where TL performs worst. 

 In the lower panels of Fig. \ref{fig3}, we show the MSE averaged over 1000 test samples.  It is evident that for the integrable model, transfer learning can achieve better precision with a lot smaller training set size in comparison with the direct training. This difference is less evident in the non-integrable case.  This is again attributed to the lower accuracy of the source model for non-integrable models.  It is worth nothing that for the non-integrable  case still the accuracy of the predictions for both TL and direct training is reasonable. 
 
In Fig. \ref{fig3}, we also show the performance of the front-end model.  As we pointed out  previously for the front-end model, LSTM layers would be more suitable and may lead to a better performance but we use dense layers here to provide a fair comparison with the trainable part of the TL model. We recall that in the TL model we choose dense layers for the trainable part. To make the comparison even more fair we also used the same number of samples for training. For the integrable model, it is evident that the TL model  predicts entropy with higher accuracy  in comparison with both direct training and the front-end model. This is another signature that confirms the usefulness of a pre-trained network. This difference is smaller for the non-integrable model. But, as we already pointed out, this is due to the low accuracy of the source model.

We would like to highlight that, in general, it is an interesting observation on its own that  while the accuracy of the source model in predicting the evolution of entropy depends on whether the model is integrable or not, this seems not to be the case for the task addressed by the front-end model. The front-end model predicts the entropy for both integrable and non-integrable models with reasonable and similar precision. We view this observation as an encouraging indication that the performance bottleneck for transfer learning in the non-integrable case is the accuracy of the source model.

Note that as demonstrated in \cite{mohseni2022deep}, our trained neural network on random Gaussian fields can extrapolate its predictions to other classes of magnetic fields, such as quench and periodic fields. Dynamics, driven by quench and periodic fields, are generally interested in quantum dynamics \cite{calabrese2011quantum, kormos2017real, navarrete2021floquet}. 

Predicting the entanglement entropy is already a fairly complex task, which, as this paper has shown, clearly can benefit from transfer learning. However, our numerical experiments have shown us that this benefit of transfer learning is not a completely generic phenomenon: indeed, we carried out preliminary investigations where we observed that even the prediction of second-order observables based on a model pre-trained on first-order observables did not yield any discernible benefits of transfer learning.


        

Our 
last remark concerns the feasibility of providing data for training the transfer learning model for large system sizes where calculating entropy numerically or measuring it experimentally is challenging. Through our numerical experiments, we have discovered the number of samples required for transfer learning on entropy is relatively small. However, it is important to acknowledge that even with this advantage, a modest amount of samples is still necessary for effective training. To overcome this challenge, one possible approach is to calculate the entropy numerically for smaller system sizes, where it remains feasible, and then leverage a pre-trained network to extrapolate the results for larger system sizes, where data might not be readily accessible. The feasibility of such extrapolations in system size has already been supported by our previous works, where we  applied a combination of LSTM and convolutional neural networks \cite{mohseni2023deep, nmohseni2021deep}.

\section{Conclusion and Outlook}
        \label{sec:conc}
        
Our findings validate that a neural network trained on the dynamics of physical observables learns useful information about the wave function. We showcase this by illustrating that the implicit representation learned by the neural network can be reused to enhance the learning of entanglement entropy. More precisely, we demonstrate that the pre-trained neural network enables more precise and resource-efficient learning of the evolution of entanglement entropy compared to direct learning. Additionally, we show that in integrable models, where the neural network exhibits superior capability in learning the physical observables, the accuracy of predictions for the evolution of entropy is also higher. More generally, our work represents a promising demonstration in the road towards exploiting representation learning and transfer learning in the context of quantum many-body dynamics.


\appendix
\section{Technical Details}
    \label{app:tech_details}
In this section, we provide a concise explanation of the neural network architectures utilized in the main text, as well as the specific subsets of observables employed to train each model.

Table \ref{tab:obs_sets}  illustrates the labels of the observables that are used to train the source model depicted  in Fig. \ref{fig2} and Fig. \ref{fig3}. Additionally, we present the architecture of the source model, including the number of hidden layers and the size of the output layer.

Regarding the target model with TL, we apply  two  fully connected trainable layers  of sizes 100 and 1 respectively.
For the target model with direct training,  two LSTM layers  of size 100 with a dense output layer of size 1 is used. 
    Note that in all cases the output layer has linear activation whereas during TL the trainable dense layers have sigmoid activation.
For the Font-end we use the same architecture as the trainable part of TL model namely two layers with size 100 and 1. 
    \begin{table}[h]
        \centering
        \begin{tabular}{l|l|l}
             label &  set & net size\\
             \hline 
             \s & $\{\langle \sigma^\alpha \rangle\}_{\alpha \in \{x, y, z\}}$ &2x\textcolor{teal}{100}+\textcolor{orange}{3}\\
             \sigsig & $\{\langle \sigma^\alpha_i\sigma^\beta_{i+l}\rangle\}_{0<l<4}^{\,\alpha,\beta \in \{x, y, z\}}$ & 2x\textcolor{teal}{500}+\textcolor{orange}{27}\\
             \Sss & $\{\langle \sigma^\alpha \rangle,\langle \sigma^\alpha_i\sigma^\beta_{i+l}\rangle\}_{
                 0<l<4}^{
                 \alpha,\beta \in \{x, y, z\}}$ & 4x\textcolor{teal}{500}+\textcolor{orange}{30}
        \end{tabular}
        \caption{Explanation of labels  used for observable sets to train the source model and the layout of the source model. Teal numbers describe the LSTM layer sizes, orange, the dense output layer.}
        \label{tab:obs_sets}
    \end{table}
    
\section{Additional considerations}\label{app:additional}
\begin{figure}[t]
    \centering
    \includegraphics[width=0.5\textwidth]{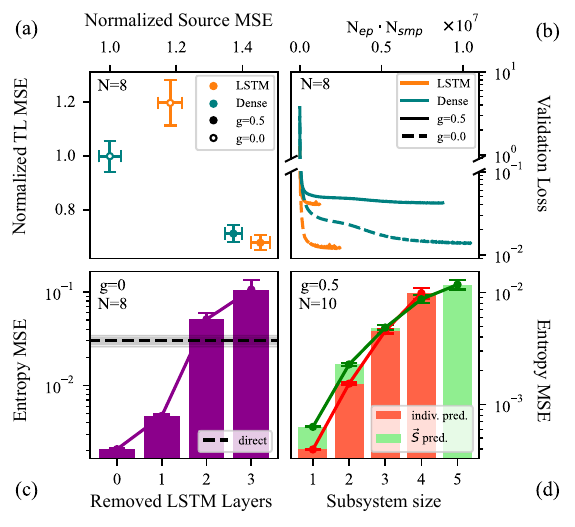}
    \caption{\textbf{(a)} LSTM vs Dense output layer in the source model. Shown results are for the case that the source model is trained  on the \s  observable set. The MSE is normalized to the MSE of the integrable model where a dense output layer is used. \textbf{(b)} Training history in TL for LSTM vs Dense trainable layers. In this case, the source model is trained on the \Sss observables  with a training set size of 5$K$. \textbf{(c)} Leveraging information from the source model to the transfer Learning. Shown is  TL performance based on discarding  different numbers of LSTM hidden layers  in the source model starting from the very last layer.  The source model is trained on the \Sss observable set. The black line is the results for the  target model for the sake of comparison. \textbf{(d)} Performance dependence of the direct trained model to the subsystem size considered to calculate entropy. The shown results are for  system size $N=10$.  Orange color shows the performance of  the direct trained model  for the case that the neural network is trained separately on entropy calculated for different subsystem sizes.  Green color shows  the neural network performance when it is trained simultaneously on a vector of entropy with its elements representing entropies calculated for different subsystem sizes. }
    \label{fig:appendix}
\end{figure}
Here we discuss some additional considerations in support of the claims made in the main text. 
\paragraph{LSTM vs Dense output layer in the source model }
Fig. \ref{fig:appendix} (a) shows the impact of using an LSTM output layer vs a dense layer in the source model. Here, the source model is trained on $\langle s \rangle$ with training set size 5$K$. Note that we normalized the MSE of the source model (TL model) to the value of the MSE of the source model  (TL model) for the case that a dense output layer is employed in the source model for the case of $g=0$.   It is evident that an LSTM output layer leads to a minor improvement in both the source model and TL performance.
\paragraph{LSTM vs Dense trainable layers in the TL model } Fig. \ref{fig:appendix} (b) shows the impact of using LSTM trainable layers  instead of dense ones during TL. In this case source model is trained on the \Sss observables  with training set size of 5$K$. Shown is the evolution of the validation loss during TL for both the integrable and non-integrable case. The LSTM seems to be beneficial just in a faster convergence with only a slight advantage in final performance. 
\paragraph{Leveraging information from the source model to TL}
Fig.  \ref{fig:appendix} (c) shows TL performance for the case that the source model is trained on \Sss. We show the MSE for cases where different number of LSTM layers are transferred  before the trainable dense layers. For the sake of comparison, the direct training results are included as well. Clearly, TL advantage can be maintained only if just the first LSTM layer is removed, but already at the cost of lower performance. 
\paragraph{Network  performance in predicting entropy across varied subsystem sizes }
In Fig.  \ref{fig:appendix} for system size $N=10$ we show the MSE in learning entropy for direct training. We explore two scenarios: (1)  the network outputs a vector with elements representing entropies calculated by tracing over different subsystem sizes, and (2)  the network is separately trained to predict entropy for each subsystem size. We observe minimal differences between these two cases; however, it is evident that it becomes progressively more challenging to predict entropy accurately as the subset size increases. This observation prompted us to adopt the maximal subset size as the choice for all the results presented in the main text.

\bibliography{paper.bib}
\end{document}